\documentclass{aa}
\usepackage{lscape,color,graphicx}
\usepackage[varg]{txfonts}

\begin{document}
   \title{Shock-triggered formation of magnetically-dominated clouds}

   \author{S. Van Loo\inst{1}, S. A. E. G. Falle\inst{2},
	T. W. Hartquist\inst{1} \and T. J. T. Moore\inst{3}}
   \authorrunning{Van Loo et al.}

   \offprints{S. Van Loo\\ \email{svenvl@ast.leeds.ac.uk}}

   \institute{School of Physics and Astronomy, University of Leeds, 
	Leeds LS2 9JT, UK \and School of Mathematics, University of Leeds,
	Leeds LS2 9JT, UK \and Astrophysics Research Institute, Liverpool
	John Moores University, Birkenhead, CH41 1LD, UK}

   \date{Received date; accepted date}

   \abstract
	{}
	{To understand the formation of a magnetically dominated molecular 
	cloud out of an atomic cloud.}
        {A thermally stable warm atomic cloud is initially in static 
	equilibrium with the surrounding hot ionised gas. A shock 
	propagating through the hot medium interacts with the cloud. We follow 
	the dynamical evolution of the cloud with a time-dependent 
	axisymmetric magnetohydrodynamic code.}
	{As a fast-mode shock propagates through the cloud, the gas behind 
	it becomes thermally unstable. The $\beta$ value of the gas also
	becomes much smaller than the initial value of order unity.
	These conditions are ideal for magnetohydrodynamic waves to produce 
	high-density clumps embedded in a rarefied warm medium. A slow-mode 
	shock follows the fast-mode shock. Behind this shock a dense shell 
	forms, which subsequently fragments. This is a primary region for the 
	formation of massive stars. Our simulations show that only 
	weak and moderate-strength shocks can form cold clouds which have 
	properties typical of giant molecular clouds.
	}
	{}
     

   \maketitle

\section{Introduction}
Observations of giant molecular clouds (GMCs) show that they are 
magnetically dominated (e.g. Crutcher \cite{C99}) with values of the 
ratio $\beta$ of thermal gas pressure to magnetic pressure of roughly 
0.04. Furthermore molecular emission lines show a non-thermal broadening
component comparable to the Alfv\'en speed (Arons \& Max \cite{AM75}). 
However, on scales larger 
than the distances between molecular clouds, the thermal and magnetic pressure
are about equal. This means that the formation mechanism of the GMCs
must reduce the value of $\beta$ considerably.

While older stellar associations are devoid of molecular gas, 
molecular clouds in the Solar neighbourhood that do not harbour any young
stars, are rare.  This suggests that the time lag between the formation of
the clouds and the stars must be short. GMCs are thus likely to be transient 
and dynamically evolving structures (e.g. Hartmann, Ballesteros-Paredes \& 
Bergin \cite{HBB01}). They can be formed by compressive motions
of gravitational and/or turbulent origin (Ballesteros-Paredes et 
al.~\cite{Betal06}).
It is thought that the formation of GMCs on scales of many tens to hundreds
of parsecs is regulated by the ram pressure from supersonic flows such as
supernova remnants, superbubbles and winds of massive stars.

Several groups (e.g. Klein et al. \cite{KMC94}; Mac Low et al. \cite{ML94};
Fragile et al.~\cite{F05}) studied the interaction of a strong shock with 
a cloud. However, these studies do not concentrate on the formation of 
molecular clouds.
Instead, they follow the destruction of the cloud and its effect on the 
properties of the interstellar medium (ISM). Lim, Falle \& Hartquist 
(\cite{LFH05}; hereafter LFH05) examined the formation of molecular clouds 
by the response of a warm atomic cloud to an increase in the pressure of the 
surrounding medium. They implicitly assumed that this pressure increase
is caused by either flow convergences or  weak shocks. Their results show that 
a process relying on pressure-driven compression and radiative cooling
leads to the formation of highly magnetically dominated regions within a 
molecular cloud consistent with observations of GMCs (Audit \& Hennebelle
\cite{AH05}).

In this paper we study the transition of a cloud, triggered by a weak shock,
from the warm atomic phase to the cold molecular one.  In Sect.~\ref{sect:model}
we specify the initial conditions of our model and summarise the 
computational details. Then, we describe the dynamical evolution of the cloud 
(Sect.~\ref{sect:evolution}) and present our results 
(Sects.~\ref{sect:results} and ~\ref{sect:Medium and large clouds}). 
Finally, we discuss these results and give our conclusions in 
Sect.~\ref{sect:conclusions}.

\section{The model}\label{sect:model}
\subsection{Initial conditions}
Like LFH05, we assume that a cloud is initially in the 
thermally stable warm phase and in pressure equilibrium with the 
surrounding hot ionised gas. The hot gas itself is not in thermal 
equilibrium as it is continuously reheated by e.g. supernova explosions
and superbubbles. Therefore, the net cooling time is long. For simplicity, 
we assume that the gas behaves adiabatically. Within the cloud, 
however, heating and radiative cooling are important.

To describe the thermal behaviour of the gas in the cloud,
we adopt a simplified piece-wise power-law cooling function and
a constant background heating rate consistent with the 
standard equilibrium phase diagram (i.e. thermal pressure versus
density) of Wolfire et al. (\cite{Wetal95}). The net heating 
rate per unit volume is given by
\[
	H = \rho[0.015 - \rho \Lambda(T)]~{\rm erg\,cm^{-3}\,s^{-1}},
\]
where 
\[
     \Lambda(T) = \left\{ 
	\begin{array}	
	{l@{\quad}r}
	3.564\times 10^{16} T^{2.12}& 0~{\rm K} \leq T < 141~{\rm K} \\
	9.1\times 10^{18} T & 141~{\rm K} \leq T < 313~{\rm K} \\
	1.14\times 10^{20} T^{0.56}& 313~{\rm K} \leq T < 6102~{\rm K} \\
	1.924\times 10^8 T^{3.67}& 6102~{\rm K} \leq T < 10^5~{\rm K} \\
	1.362\times 10^{29} T^{-0.5}& T \geq 10^5~{\rm K} \\
	\end{array} \right.
\]
(S\'anchez-Salcedo et al.~\cite{SVG02}; LFH05). 
For this thermal model, the warm phase does not exist for a number density and 
thermal pressure higher than $0.5~{\rm cm^{-3}}$ and $3051k$, respectively.
Here, $k$ is the Boltzmann constant.  

In our calculations, a cloud with initial radius $R_{cl} = 200~{\rm pc}$ 
has a number density of $0.45~{\rm cm^{-3}}$. As the cloud is in the warm 
phase, its pressure and temperature, respectively, correspond to 2825~$k$ and 
6277~K. The surrounding hot gas has a number density of $0.01~{\rm cm^{-3}}$.
(Note that these values are the same as in LFH05.)
To ensure pressure equilibrium between the cloud and its 
surroundings, the temperature of the external medium is set to
282\,500~K. The initial magnetic field is uniform and in the axial 
direction. It has a value such that the ratio of thermal pressure to 
magnetic pressure is unity inside the cloud.

While LFH05 examined a cloud subjected 
to a sudden increase in the external pressure, we consider a steady, 
planar shock hitting the quiescent cloud. This is known as 
the small-cloud approximation (see Klein et al.~\cite{KMC94}). 
This intercloud shock propagates along the magnetic field lines through 
the hot ionised gas at a speed $v_{ext}$ (quantified by its 
Mach number $M \equiv v_{ext}/c_s$, where $c_s$ is the nonmagnetic 
adiabatic sound speed). Table~\ref{tab:model} lists the shock 
Mach number for each model. We include models D and E in order to
examine how deviations from the small-cloud approximation
affect the results.

\begin{table}
\caption{Shock Mach Number for the numerical models}
\label{tab:model}
\centering
\begin{tabular}{cc}
\hline \hline
Model& Shock Mach Number \\
\hline
A & 2.5 \\
B & 1.5  \\
C & 5.0 \\
D$^a$ & 2.5 \\
E$^a$ & 2.5 \\
\hline
\end{tabular}

$^a$ Medium and large cloud models.\\ More details are given in 
Sect.~\ref{sect:Medium and large clouds}.
\end{table}

\subsection{Computational details}
All of our calculations were done with an adaptive mesh refinement code
solving the ideal magnetohydrodynamic (MHD) equations for an axisymmetric 
geometry (Falle \& Giddings~\cite{FG93}). The code uses an hierarchy of
$n$ levels such that the mesh spacing is $\Delta x/ 2^n$, where $\Delta x$ 
is the spacing of the coarsest grid. The code refines the grid only 
where higher resolution is necessary. The basic algorithm is 
a second-order Godunov scheme (Falle~\cite{F91}) with a linear 
Riemann solver and a divergence cleaning algorithm (Dedner et al.~\cite{D02}). 

The computational domain is $0 \leq r/R_{cl} \leq 2$ and 
$-2 \leq z/R_{cl} \leq 3$.
The shock propagates in the negative $z$-direction and we impose 
fixed boundary conditions, i.e. the shock's downstream values, 
on the $z/R_{cl} = 3$ plane. The other boundary conditions are symmetry on the 
axis and free flow everywhere else. There are six levels of grids 
with the finest being 1280 $\times$ 3200 which gives a mesh spacing 
of 0.31 pc for our models.
Although this resolution is high enough to show the onset of the formation 
of dense clumps in the interstellar cloud, it is insufficient to follow 
the complete evolution of the cloud.

\section{Dynamical evolution}\label{sect:evolution}
Although the dynamical evolution of the cloud depends 
on the initial shock Mach number, it is similar for all our models.
The results differ only in detail. Therefore, it is 
useful to describe the evolution before presenting the 
numerical results in Sects.~\ref{sect:results} 
and~\ref{sect:Medium and large clouds}

\subsection{The flow outside the cloud}\label{sect:outside}
When the intercloud shock first hits the cloud, it transmits  a
shock into the cloud and reflects a shock in the previously shocked 
intercloud medium. The reflected shock eventually forms a bow shock 
(or bow wave) in front of the cloud, reducing the flow speed of
the incoming gas. 
After a shock-crossing time $t_{sc} = 2R_{cl}/v_{ext}$, the intercloud
shock has swept around the cloud and converged on the axis behind it. 
Since the curved, converging 
part of the intercloud shock is moving at a lower speed than the 
straight part, swept-up material is raised to a lower pressure. 
The transmitted shock is, therefore, weaker at the sides of the cloud 
than on the front and back. As a consequence, the shock mainly reduces 
the cloud size in the direction parallel to the distant upstream flow 
velocity making it oblate. Eventually,
the cloud ends up as a thin disk. The ram pressure of the external 
flow at the front of the cloud even accelerates this process.

The transmitted shock has a characteristic speed $v_{int} = v_{ext}/\chi^{1/2}$ 
where $\chi$ is the density ratio of cloud/intercloud gas. This is considerably
lower than the propagation speed of the intercloud shock. Hence, a 
 velocity shear layer forms at the edge of the cloud. This slip surface
rolls up to produce a vortex ring sweeping material away from 
the cloud and is also  subject to Kelvin-Helmholtz instability 
(Klein et al.~\cite{KMC94}). However, linear instability analysis 
(e.g. Chandrasekhar \cite{C61}) predicts that a magnetic field strongly 
suppresses the onset of Kelvin-Helmholtz instability. This is also 
the case for Rayleigh-Taylor instability that is excited at the cloud 
surface due to the acceleration of the cloud by the rarefied surrounding gas. 
Although the inhibition of these instabilities limits the 
disintegration of the cloud (as seen in hydrodynamic simulations),  
some fragments still get torn off which survive as small-scale coherent 
structures (MacLow et al.~\cite{ML94}).

Because the external flow is diverted around the cloud, the magnetic 
field structure changes significantly. When the shock engulfs the cloud,
the intercloud gas near the most upstream point of contact with 
the cloud is swept 
around the cloud and carries the magnetic field with it. Since the magnetic 
field lines are anchored in the cloud, they are stretched into a self-reversed
layer (Mac Low et al.~\cite{ML94}). Eventually, these lines double over 
and form a region of 
strong reversed field. This region may be subject
to magnetic reconnection. 

\subsection{The flow inside the cloud}\label{sect:inside}
A fast-mode shock and a trailing slow-mode shock are driven
into the cloud. The fast-mode shock compresses the gas so 
that it is in the thermally unstable temperature range. 
As the time scale for the radiative cooling is much shorter than
the time for the shock to be reflected at the axis, given by 
the cloud-crushing time $t_{cc} = R_{cl}/v_{int}$,
the gas loses a significant fraction of its internal 
energy during the compression. This initiates the formation of cold dense 
regions embedded in a diffuse warm medium. These clumps
continue to cool and accumulate material until the cooling balances 
the heating. The clumps end up in approximate 
\begin{landscape}
\begin{figure}
\begin{center}
\resizebox{7.3cm}{!}{\includegraphics{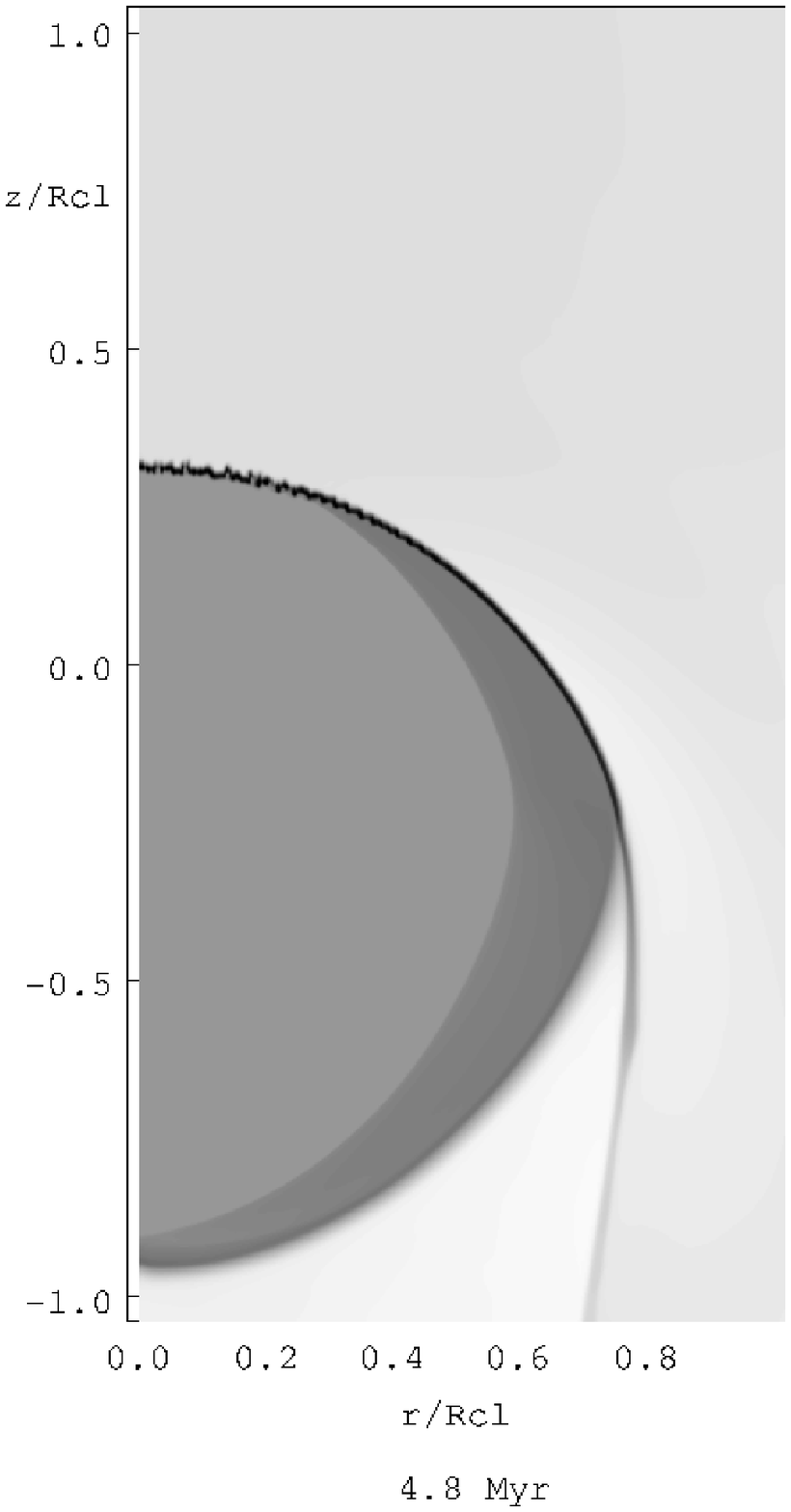}}
\resizebox{7.3cm}{!}{\includegraphics{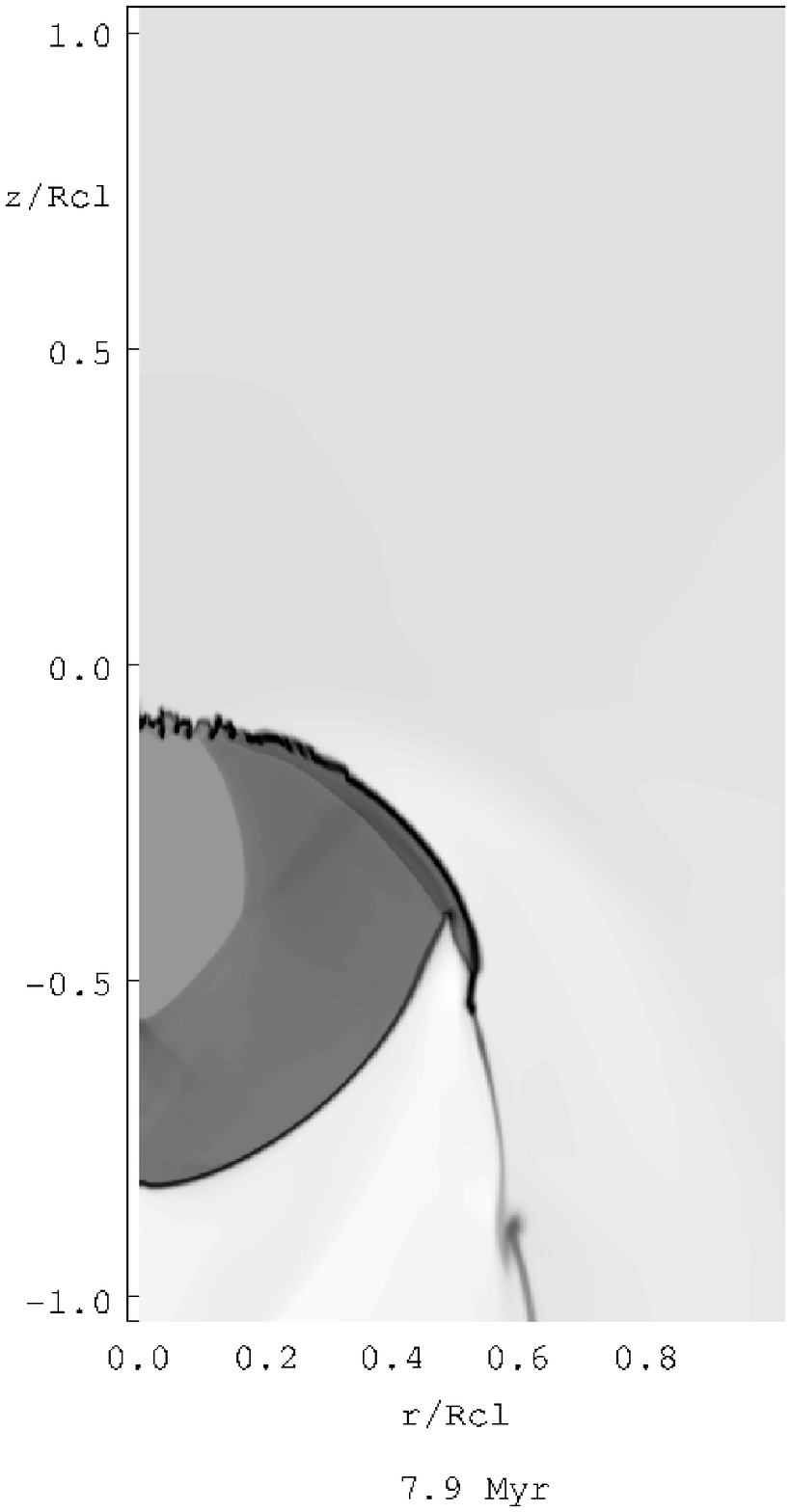}}
\resizebox{7.3cm}{!}{\includegraphics{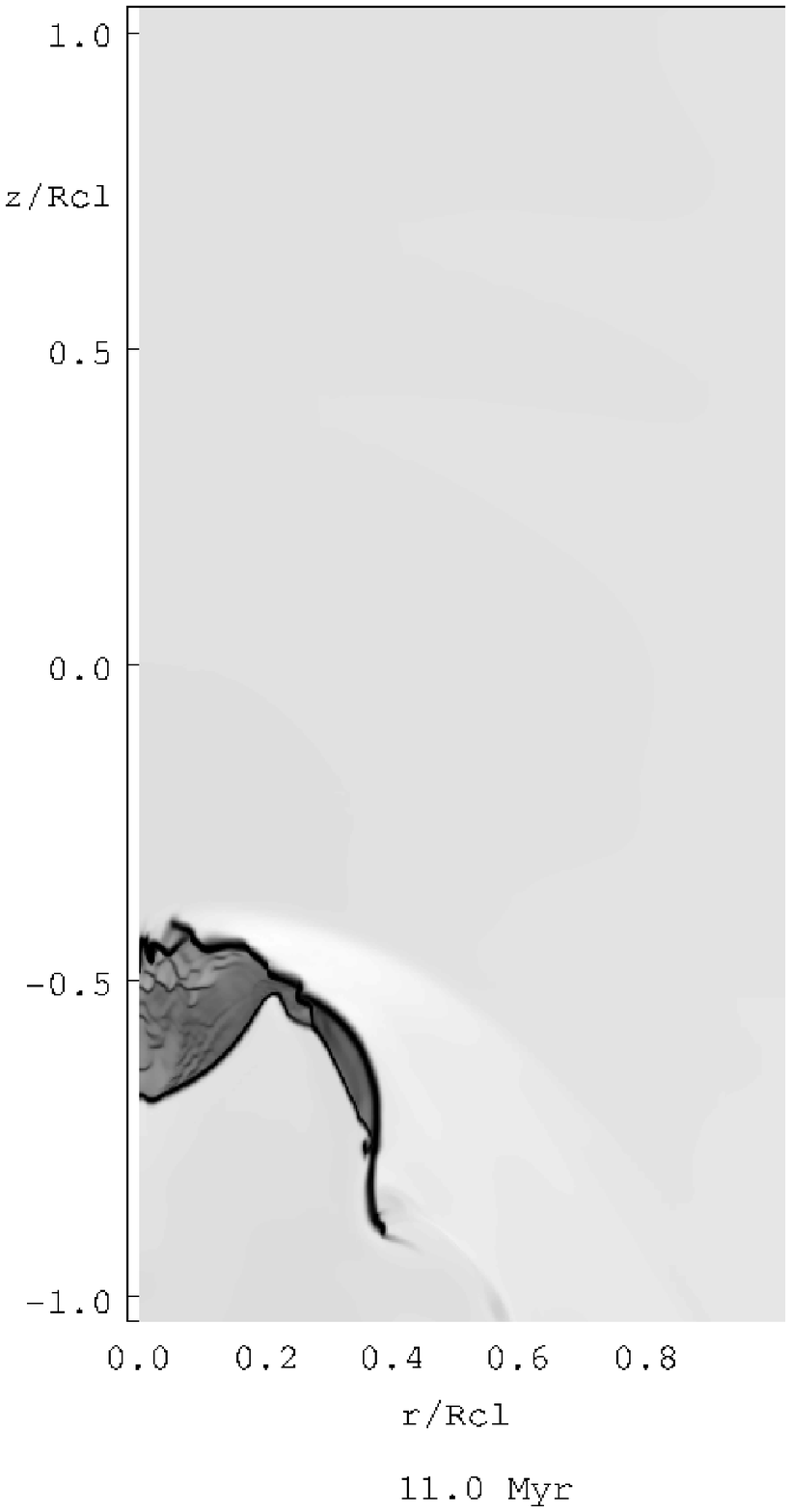}}
\hspace{0.5cm}
\resizebox{1.68cm}{!}{\includegraphics{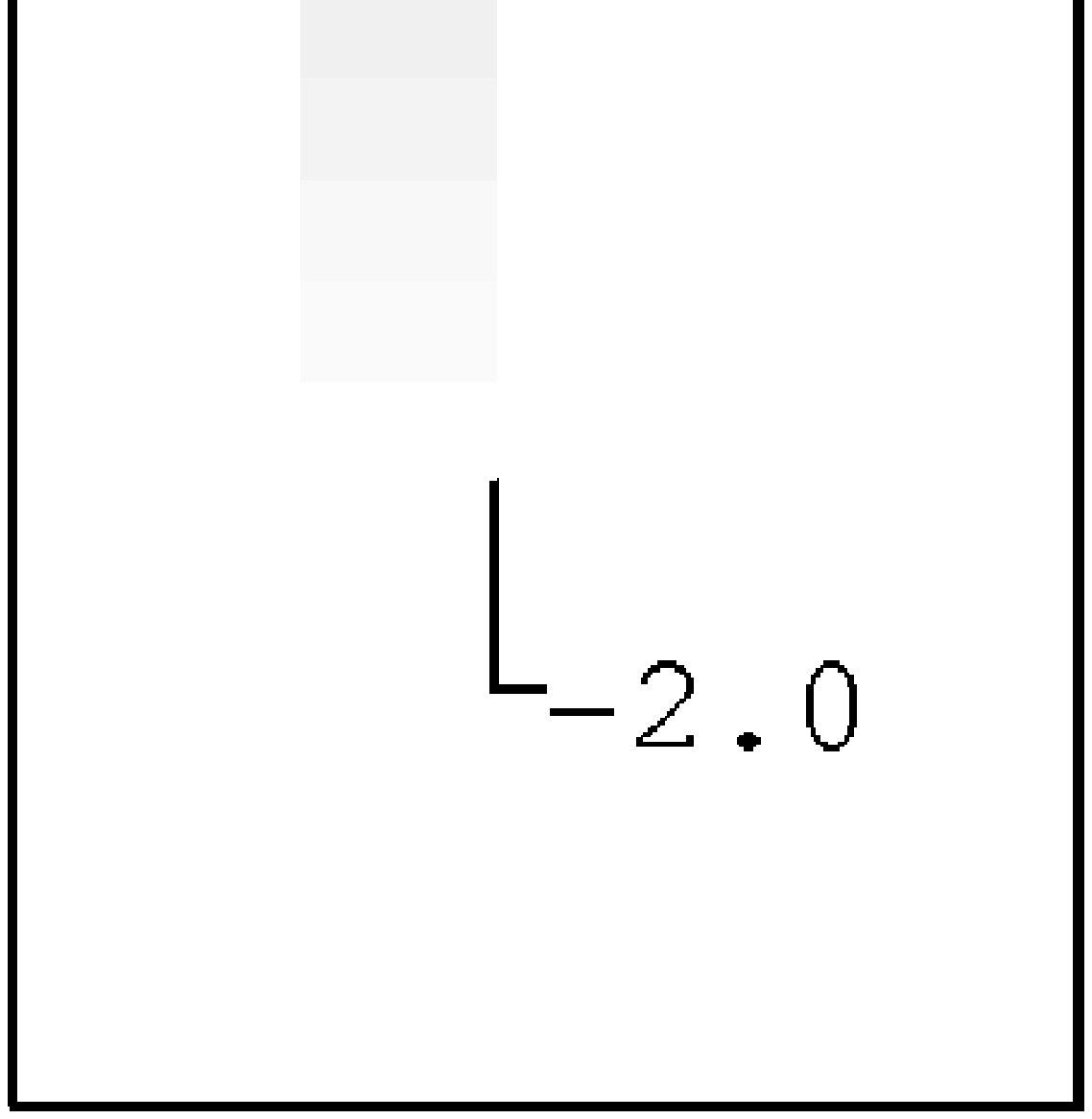}}\\
\end{center}
\caption{Grey-scale plots of the number density (logarithmic, {\it a})
and $\beta$ (linear, {\it b}) at the times shown for model A. 
For the number density the range is -2 -- 2 and for $\beta$ it is 0 -- 1.
The size of the computational box shown is $0 \leq r/R_{cl} \leq 1$ and 
$-1 \leq z/R_{cl} \leq 1$. The post-shock flow is in the 
negative $z$-direction.}
\label{fig:evolution}
\end{figure} 
\end{landscape}
\setcounter{figure}{0}
\begin{landscape}
\begin{figure}
\begin{center}
\resizebox{7.3cm}{!}{\includegraphics{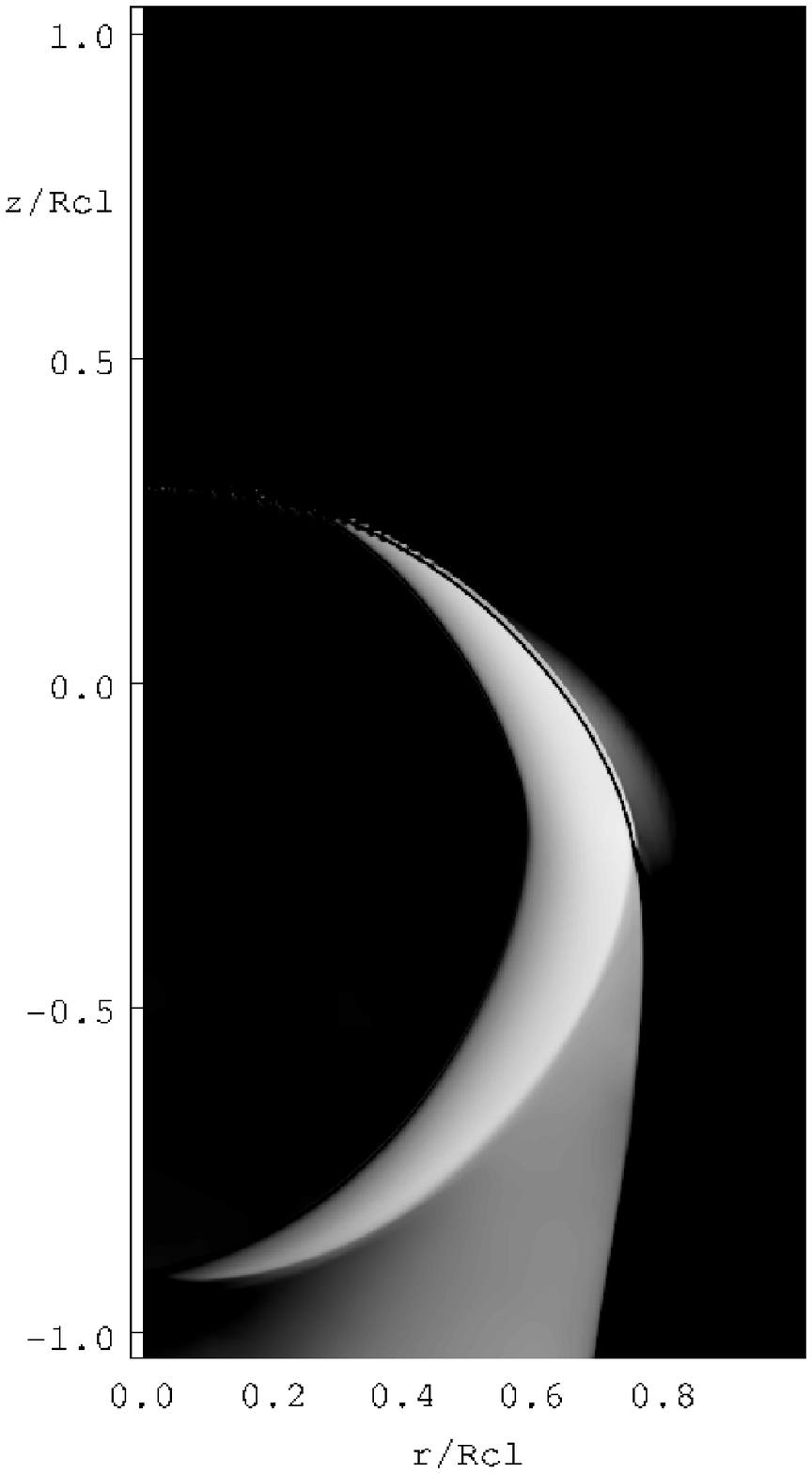}}
\resizebox{7.3cm}{!}{\includegraphics{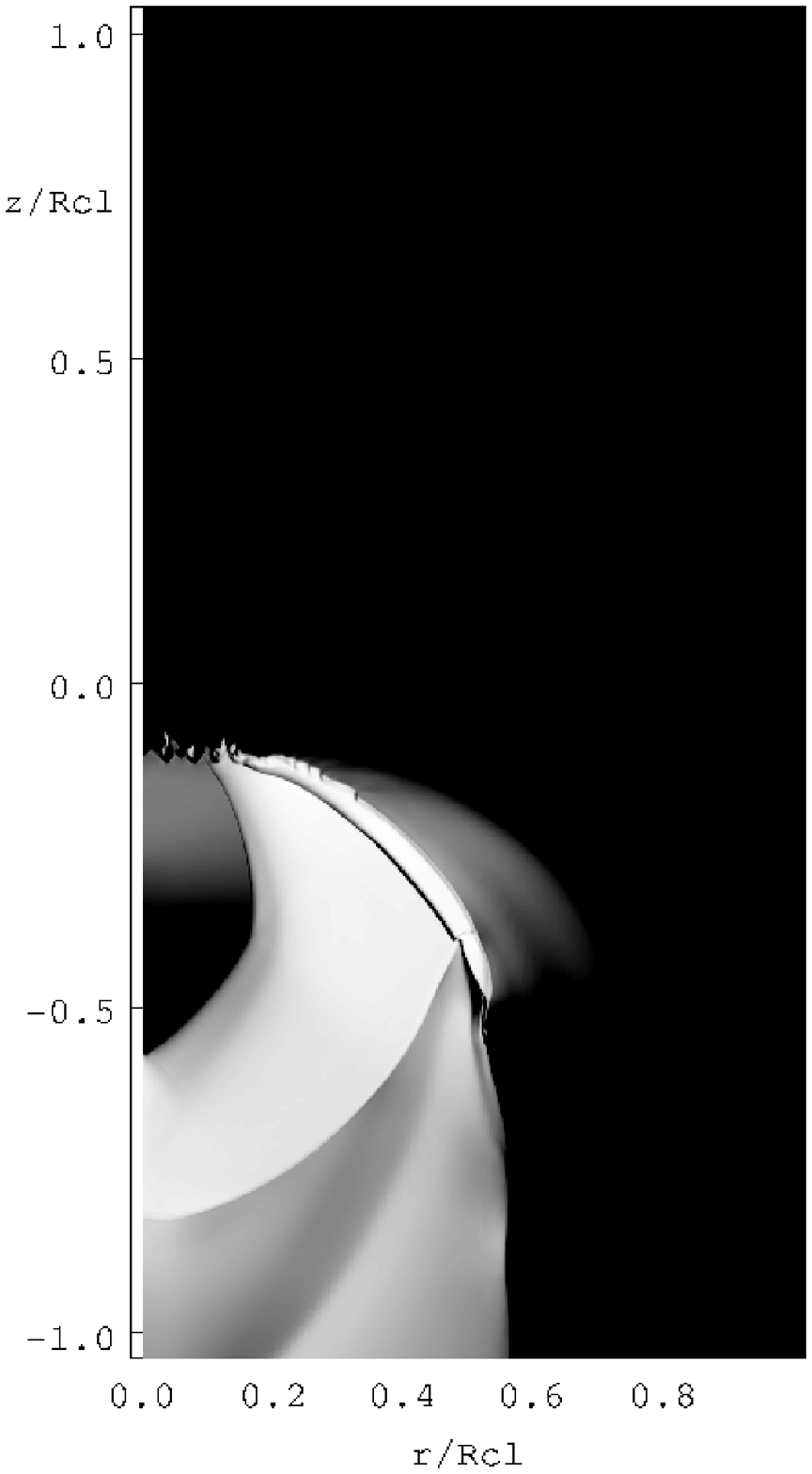}}
\resizebox{7.3cm}{!}{\includegraphics{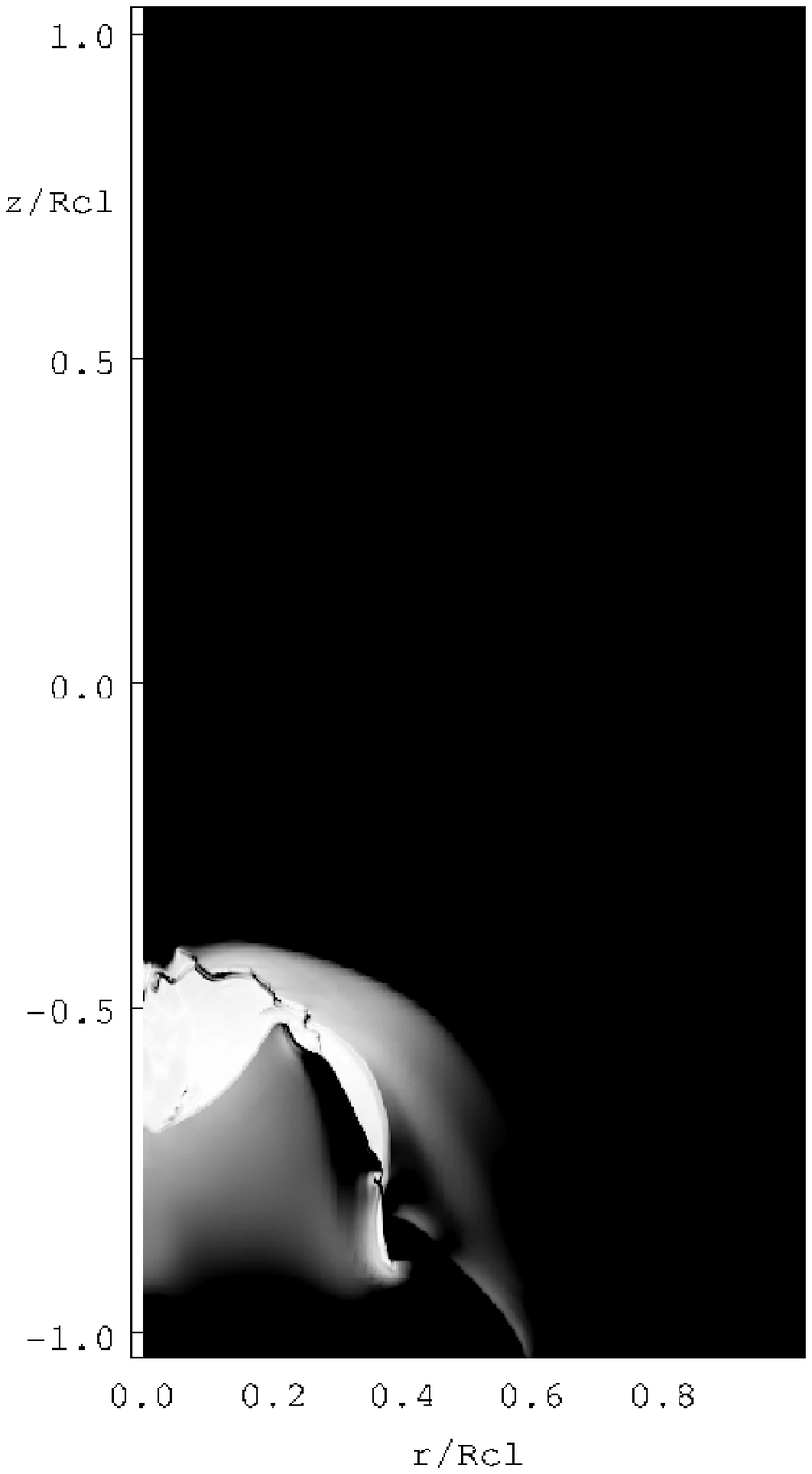}}
\hspace{0.5cm}
\resizebox{1.70cm}{!}{\includegraphics{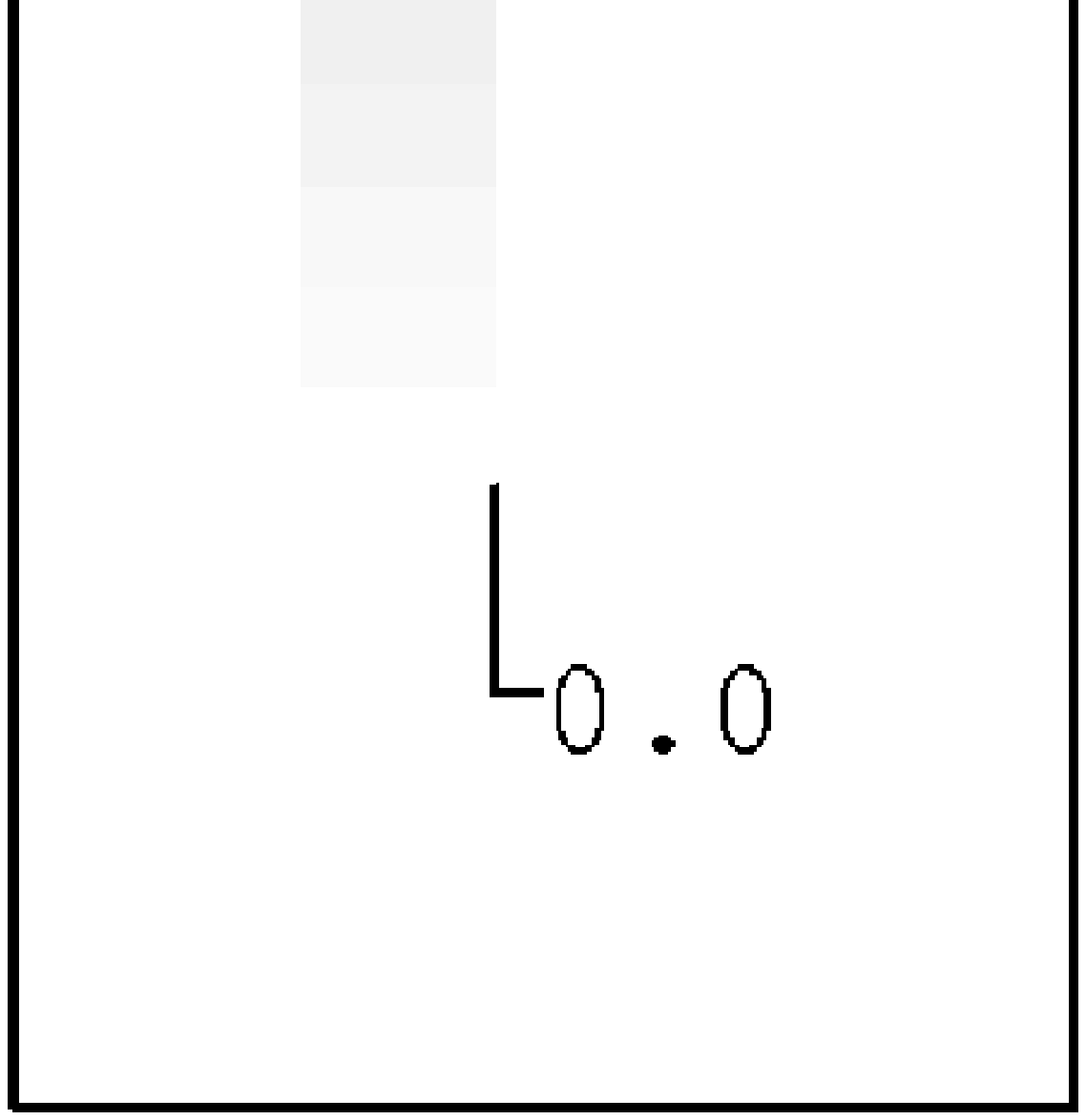}}\\
\end{center}
\caption{(\it continued)}
\end{figure} 
\end{landscape}
pressure equilibrium with their surroundings. The number density of a clump is 
then in the range 500 - 1\,000 cm$^{-3}$ which is typical for 
translucent clumps in GMCs (e.g. Williams et al.~\cite{Williamsetal95}). 

Furthermore, as the magnetic pressure increases across the fast-mode shock,
the value of $\beta$ falls below unity. This provides the ideal conditions
for the generation, by MHD waves, of dense cores within these clumps
(e.g. Falle \& Hartquist \cite{FH02}; Van Loo, Falle \& Hartquist~\cite{VFH06}),
especially as thermal instability produces a supersonic velocity 
dispersion within the cloud (Koyama \& Inutsuka \cite{KI02}). However, 
we do not have sufficient spatial resolution to follow the complete 
clump and dense core formation process. Moreover, self-gravity, which is 
not included in our models, becomes dynamically important once some dense
cores are compressed to their measured number densities.    

The slow-mode shock moves much more slowly 
than the fast-mode shock. The magnetic pressure decreases behind this  
shock, and consequently there is nothing to stop the gas from 
compressing while it cools. Hence, a shell of cold, dense gas forms 
close to the edge of the cloud. The dense shell first
develops on the upstream parts of the cloud, even before the shock 
has interacted 
with the rest of the cloud. Furthermore, the velocity shear at the surface 
of the cloud leads to a broader dense shell near the stagnation point
(see e.g.~Fig.~\ref{fig:evolution}) as the gas compressed by the slow-mode 
shock is swept around the cloud. The thickness of this shell 
thus depends on the level of shear at the edge, i.e. the shock Mach number.

The interaction of the shock with the cloud produces dense regions in 
the cloud, i.e. a shell at the edge of the cloud and several
individual clumps within the cloud. The shell contains a large 
fraction of the cloud's total mass. It  is also unstable due to thermal 
instability and dynamical instabilities (such as Kelvin-Helmholtz and 
Rayleigh-Taylor), so that it fragments into cold, massive filaments. 
Our simulations show this fragmentation at later times. This means 
that the shell may well be the precursor of massive star formation. The 
clumps within the cloud, may form only low-mass stars as they are magnetically
dominated.

\section{Numerical results}\label{sect:results}
\begin{figure}
\resizebox{\hsize}{!}{\includegraphics{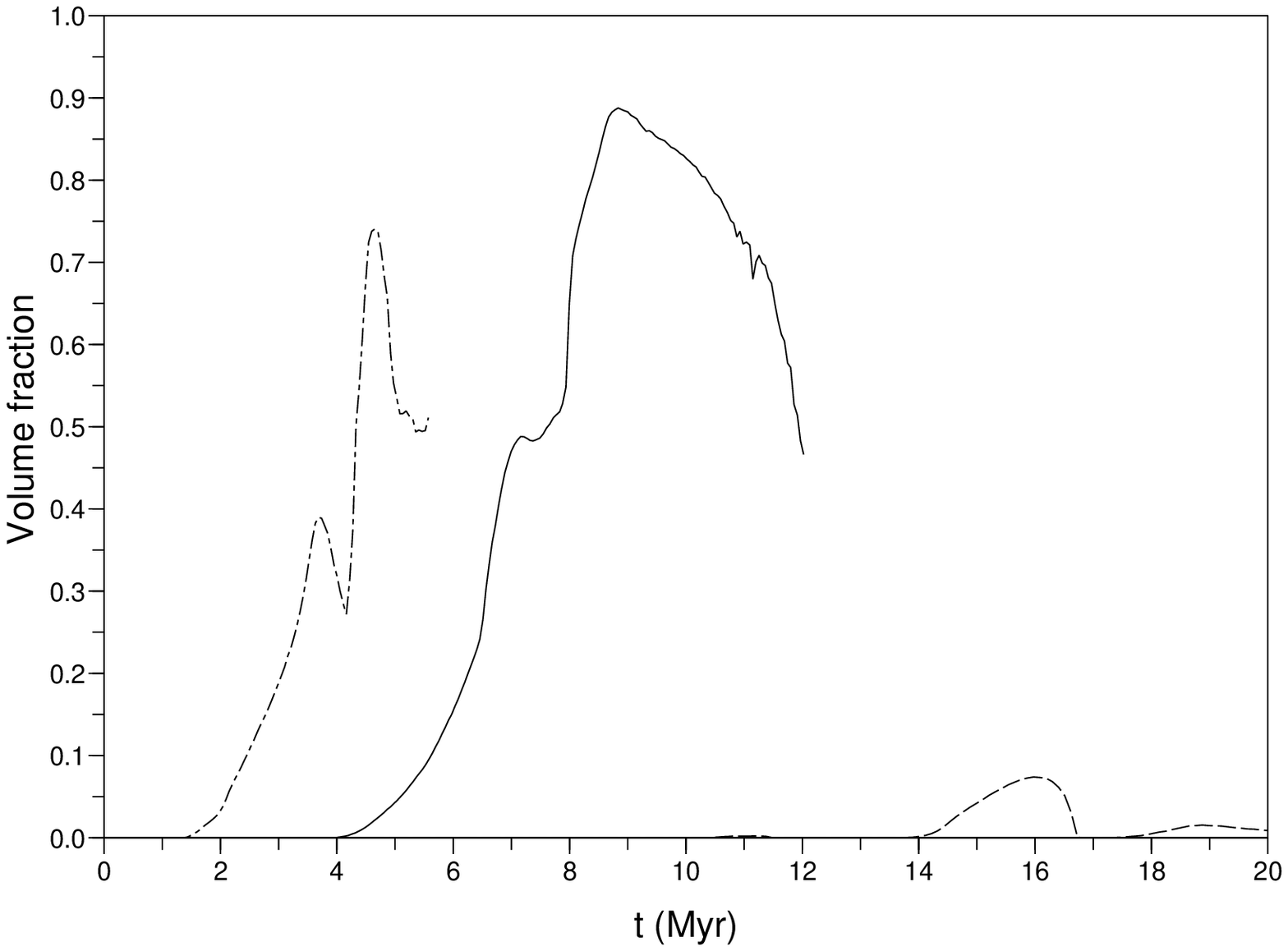}}
\caption{The fraction of the cloud volume with $\beta < 0.1$ as a function
of time (in Myr) for Model A (solid), Model B (dashed) and Model C 
(dashed-dotted). }
\label{fig:volumefraction1}
\end{figure}

Figure~\ref{fig:evolution} shows grey-scale plots of the number density and 
$\beta$ for model A. We can clearly see that, as the fast-mode shock moves 
through the cloud, the swept-up gas becomes magnetically dominated, i.e. 
$\beta$ drops below unity. This is because the thermal gas pressure decreases 
due to the transition to the thermally unstable phase, while the magnetic 
pressure increases behind the fast-mode shock.
It can then be expected that, by the time the shock reflects at 
the symmetry axis, a significant fraction of the volume of the cloud has 
a low-$\beta$ value. 
Figure~\ref{fig:volumefraction1} shows the fraction of cloud volume 
with $\beta < 0.1$ as a function of time for each of three models. 
At around $t_{cc}$, the volume fraction of the model A cloud for which 
$\beta < 0.1$ is about $\sim 0.9$.
Actually, most of the cloud is in a low-$\beta$ regime for a significant 
period of time ($\sim 3$~Myr). Within this time range, the cloud has 
a characteristic mean number density and radius of  $\approx 20~{\rm cm^{-3}}$ 
and $\approx 50~{\rm pc}$.  Furthermore, we find that the velocity dispersion 
of the cloud is highly supersonic. These are all typical properties of a GMC.

\begin{figure}
\resizebox{\hsize}{!}{\includegraphics{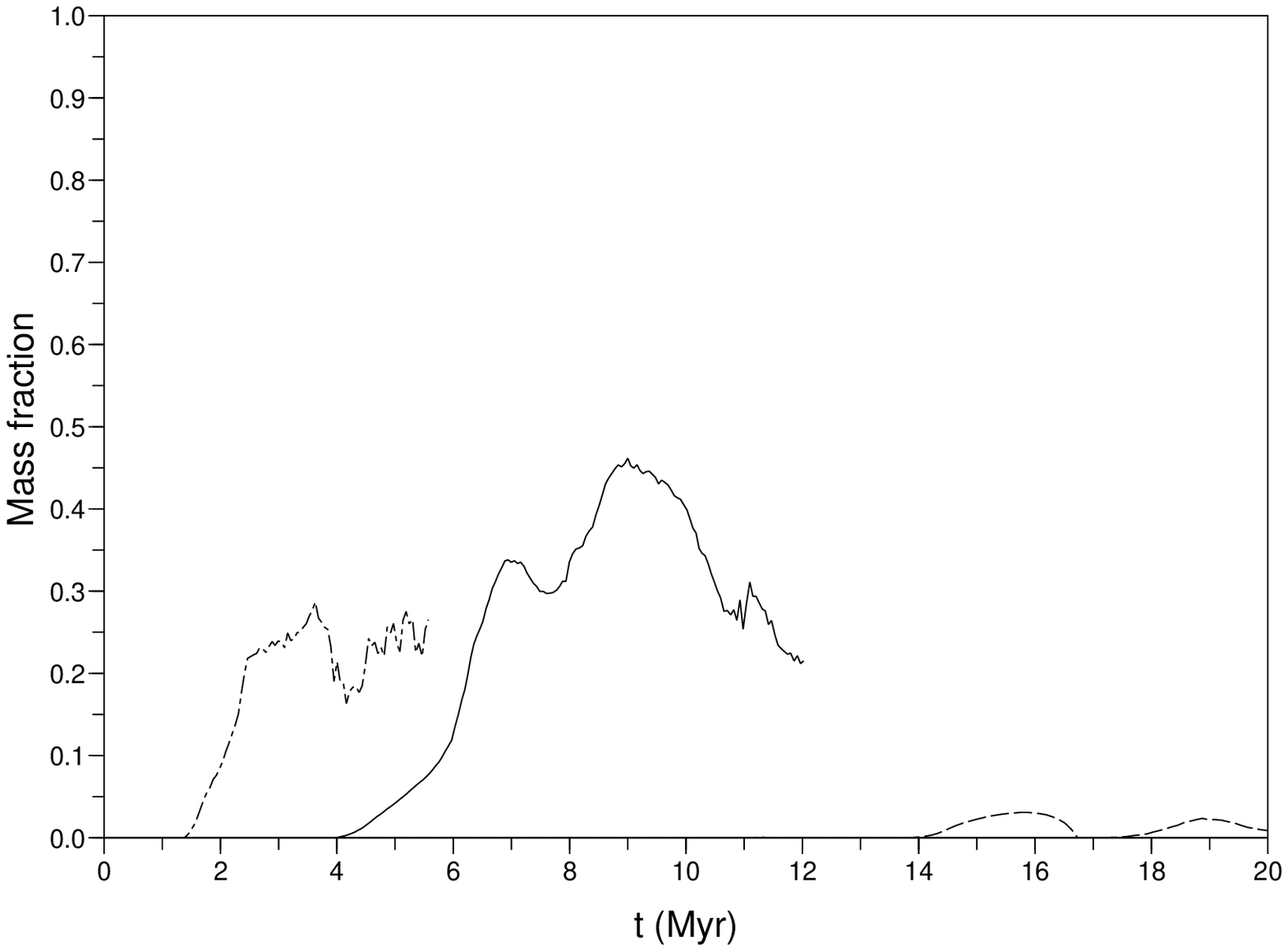}}
\caption{The fraction of the cloud mass with $\beta < 0.1$ as a function
of time (in Myr) for Model A (solid), Model B (dashed) and Model C 
(dashed-dotted).}
\label{fig:massfraction1}
\end{figure}

Although a small fraction of the volume of cloud is not magnetically dominated,
its corresponding mass fraction is considerable (see 
Fig.~\ref{fig:massfraction1}). About half of the total mass is 
in the boundary layer of the cloud. Here, the gas is in the  
cold phase at a high thermal pressure so that $\beta \approx 1$.
While the conditions in the interior of the cloud are ideal for the 
generation of 
dense clumps and embedded cores by MHD waves, this is not true for 
the dense shell. It is, however, expected that this shell breaks up into 
dense fragments (Koyama \& Inutsuka \cite{KI00}). This fragmentation is 
observed in our simulations. The densest fragments have number densities of 
$\sim 10^3$~cm$^{-3}$ and sizes of $\sim$~1 pc. 
With a temperature of about 30~K for these 
fragments, we derive a Jeans mass of $\sim 350~{\rm M_{\sun}}$ which is 
similar to the inferred mass. (We have assumed a near spherical shape for the 
dense fragments.) Therefore, the boundary layer may be
a site of massive star formation. 

For the other models, we find similar results, except that stronger shocks
(Model C) lead to a faster evolution (see Fig.~\ref{fig:volumefraction1}). 
This is because the propagation speed, $v_{int}$, of the shock increases with 
Mach number, reducing the dynamical time-scale $t_{cc}$. (The cloud
crushing time is $t_{cc} \approx 4~{\rm Myr}$ for Model C, while 
$t_{cc} \approx 8.5~{\rm Myr}$ for Model A.) Magnetically-dominated regions
thus develop on shorter time-scales. However, the compression of the cloud into
a disk also proceeds faster due to a higher ram pressure exerted by 
the external flow (i.e. the ram pressure is proportional to $M^2$). 
Our simulations show that, for Model C, the 
time-scale for flattening is $\sim 1 t_{cc}$, while it increases 
to $\sim 1.5 - 2 t_{cc}$ for weaker shocks (Model A and B). 
Consequently, the lifetime of a molecular cloud produced by a strong 
shock has an upper limit of a few Myr before it is flattened. 
The formation of H$_2$ and CO from atomic gas requires a similar timescale 
(Bergin et al.~\cite{Betal04}; Glover \& Mac Low \cite{GM07}). 
Thus, as the dynamical evolution is too rapid for hydrogen to be completely
converted to H$_2$, strong shocks do not produce giant molecular clouds.
However, molecules can still form in the flattened
cloud. As this thin disk is being shredded by instabilities (MacLow et al.
\cite{ML94}), this results in the formation of small molecular clouds 
with sizes of a few pc.  

The time-scale for molecular cloud formation by a weak shock (Model B)
is compatible with the expected time-scales for the formation of 
molecules (the cloud flattens on a time-scale of $\sim 25$ Myr). 
However, in this case the fast-mode shock propagating through the cloud 
does not compress
the gas sufficiently to induce fast radiative cooling. As the cooling time
of the gas is longer, it is
difficult to produce magnetically-dominated regions within the cloud. 
Figure~\ref{fig:volumefraction1} shows that less than 10 \% of the 
cloud has $\beta < 0.1$ during its entire evolution. Note, however, 
that we have not included any feedback effect of star formation.
It is believed to take place in the dense boundary layer of the cloud. 
We can then expect proto-stellar outflows to reduce the value of $\beta$
locally, thus, forming magnetically dominated regions.

\section{Medium and large clouds}
\label{sect:Medium and large clouds}

In the previous section we showed the results for a shock, being driven by
a constant distant upstream pressure, interacting
with a cloud; i.e. the so-called small-cloud approximation. However, the 
pressure behind a wind-blown bubble or a supernova blast wave can change 
significantly 
on a time-scale shorter than $t_{cc}$, or even $t_{sc}$. By comparing 
these time-scales to the time-scale, $t_{p}$, of pressure 
variations we can make a further distinction between {\it medium-sized} and 
{\it large} clouds (Klein et al.~\cite{KMC94}). For medium-sized clouds, 
the pressure
does not vary much as the shock sweeps around the cloud, but it does change
significantly before the cloud is crushed, i.e. $t_{cc} \geq t_{p} \geq t_{sc}$.
For large clouds, $t_{p} < t_{sc}$, and the pressure thus changes while 
the shock sweeps around the cloud. 

To study the effect of pressure variation behind the shock, we assumed that 
a shock is accelerated exponentially from M = 1.1 to a higher value of 
$M$ (given in Table~\ref{tab:model} for models D and E) after which it 
accelerates no more. In our calculations, we assumed that the period of 
constant shock strength is $\sim$ 1~Myr for a large
cloud (Model E) and $\sim$ 5 Myr for a medium-sized cloud (Model D). 
Our model is a simplified description of a blast wave. However, we prefer 
this to the implementation of the Sedov-Taylor blast-wave solution, 
as we are only
interested in the qualitative differences from the small-cloud approximation. 

\begin{figure}
\resizebox{\hsize}{!}{\includegraphics{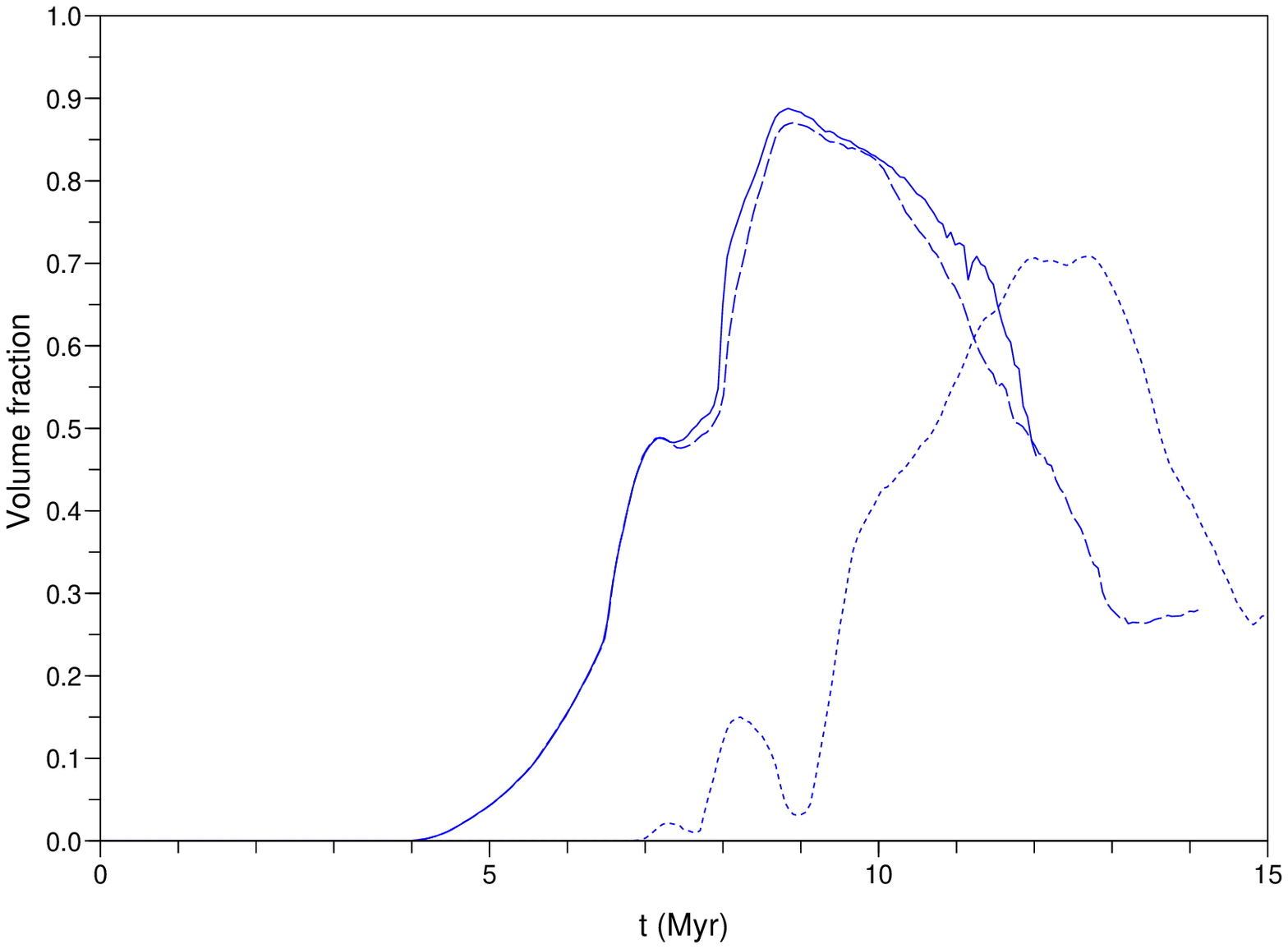}}
\caption{Similar to Fig.~\ref{fig:volumefraction1}, but now for small (solid),
medium (dashed) and large (dotted) cloud approximation.}
\label{fig:volumefraction2}
\end{figure} 

Figure~\ref{fig:volumefraction2} shows that, for medium-sized clouds, 
the results
do not change significantly from the small-cloud approximation. Large clouds,
however, evolve at a lower rate towards  a low-$\beta$ regime.
This can be readily explained by the fact that 
the pressure changes as the shock sweeps around the cloud so that
the shock strength on the upstream and downstream sides is quite different.
The fast-mode shock therefore does not produce the same compression of 
the gas as in the small-cloud approximation. This results in a longer 
cooling time and a delay in the formation of a low-$\beta$ cloud.  
Yet, most of the cloud in the large-cloud approximation still becomes 
magnetically dominated.

\begin{figure}
\hspace{-0.3cm}
\resizebox{4.6cm}{!}{\includegraphics{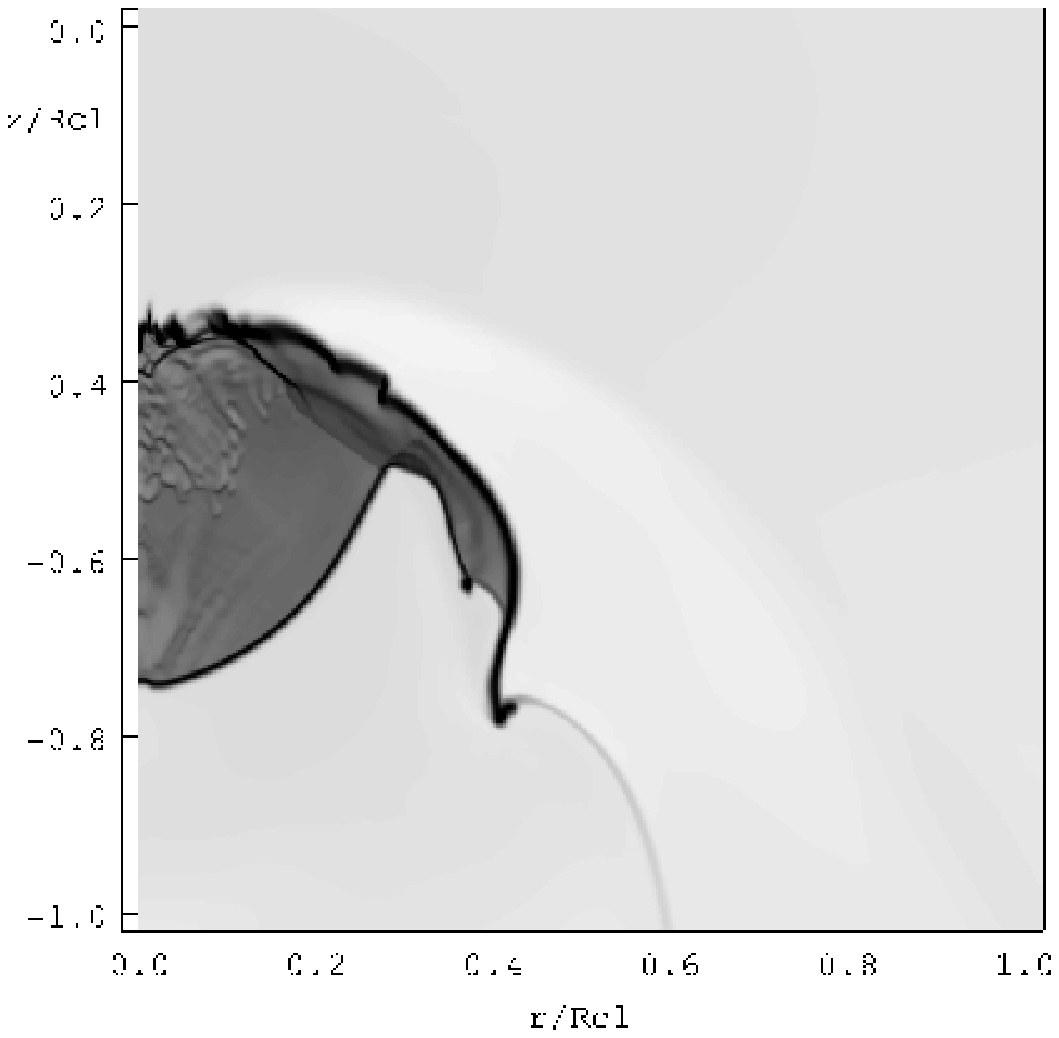}}
\hspace{-0.4cm}
\resizebox{4.6cm}{!}{\includegraphics{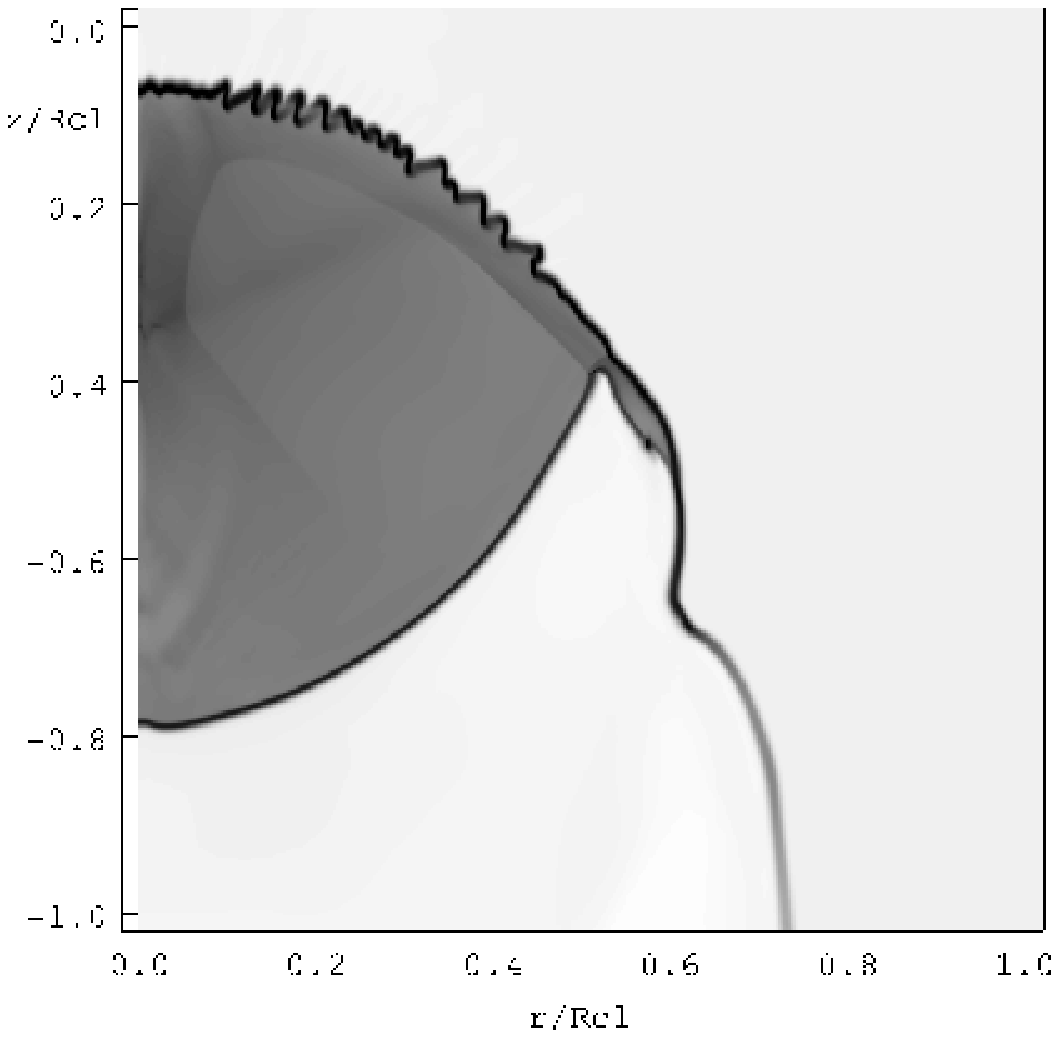}}
\caption{Logarithmic grey-scale plot of the number density for Model A 
({\it left}) and Model E ({\it right}) at 10~Myr. The range for the 
number density is -2 -- 2.}
\label{fig:compression}
\end{figure}

The fact that the pressure behind the shock decreases on time-scales 
shorter than $t_{sc}$ has the additional advantage that the cloud is 
exposed to a lower ram pressure. Consequently, the cloud does not 
flatten into a disk as fast as in the small-cloud approximation
(see Fig.~\ref{fig:compression}). This is relevant as it shows 
that shocks, stronger than the small-cloud approximation would indicate,
can produce clouds resembling GMCs. 

\section{Conclusions and discussion}\label{sect:conclusions}
In the present paper we studied the interaction of a shock with 
an initially warm, thermally stable cloud of uniform density. 
As the shock sweeps around
the cloud, a transmitted fast-mode shock propagates through the cloud
producing magnetically-dominated regions behind it. This provides 
the ideal conditions for MHD waves to produce high-density clumps within
the cloud. These clumps may eventually form low-mass stars.

A slow-mode shock follows the fast-mode shock leading to a high-density 
layer near the boundary of the cloud. As the shock interacts first with the 
front of the cloud, the dense layer forms there first. The dense layer is 
subject to instabilities and fragments into massive clumps. Our analysis 
shows that these clumps are close to gravitationally bound and, thus, 
are possible precursors of massive stars. This agrees well with the findings 
that OB stars form almost always near the boundary of molecular clouds 
and even prefer to form at one particular edge of the cloud (e.g. 
Blaauw \cite{B62}; Elmegreen \& Lada \cite{EL77}; Israel \cite{I78}; 
Gatley et al.~\cite{G79}; Fich et al.~\cite{F82}). 
Furthermore, the cometary cloud structure seen in our simulations, i.e. 
a massive head and long-spread tail, seems characteristic for clouds
harbouring cluster-forming cores (Tachihara et al.~\cite{T02}).

In our simulations we imposed a sharp boundary between the cloud
and the surrounding medium. However, real interstellar clouds have an 
internal density gradient, e.g. the southwest cloud in the Cygnus Loop
supernova remnant 
shows a gradual drop in density near the boundary (Patnaude et al.~\cite{P02}).
Hydrodynamical simulations of the effect of smooth boundaries show that 
the initial density pattern remains imprinted on the shocked cloud
(Nakamura et al.~\cite{Netal06}).  Preliminary results of MHD 
simulations show a similar effect, but we still find that a thin dense layer 
arises near the boundary of the cloud. The effect of a density gradient 
and the presence of condensations within the cloud will be investigated 
in detail in a subsequent paper. 

\begin{figure}
\resizebox{\hsize}{!}{\includegraphics{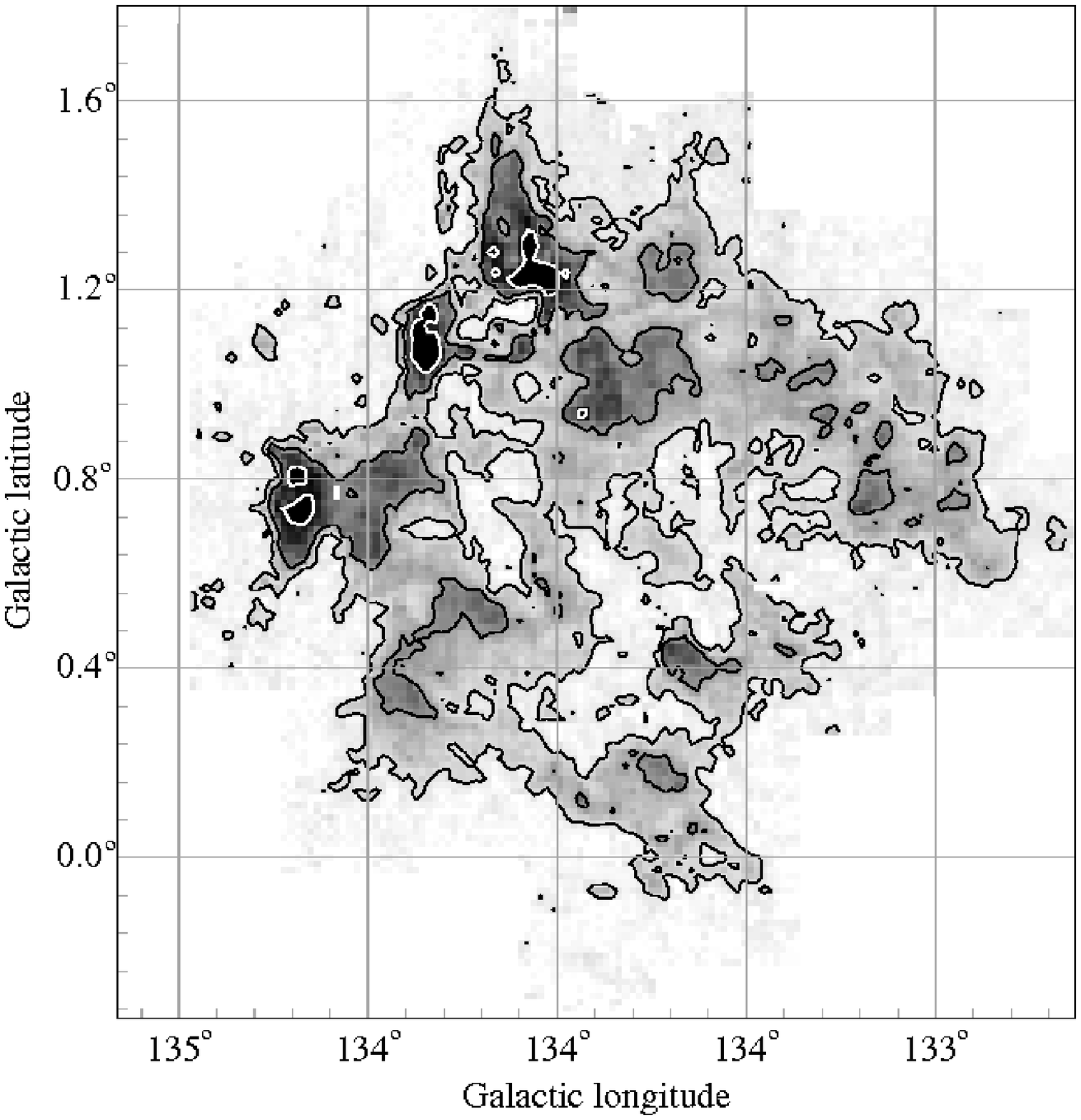}}
\caption{The W3 GMC in the J=1-0 transition of $^{12}$CO with 
a 50 arcsecond resolution (Bretherton \cite{B03}). High-mass 
star-forming regions are found in the HDL situated in the North-East
of the cloud.  }
\label{fig:w3}
\end{figure} 

A likely example of a shock-cloud interaction is found in the W3 GMC.
Figure~\ref{fig:w3} shows the W3 GMC as traced by $^{12}$CO J=1-0
emission at 50 arcseconds resolution (Bretherton~\cite{B03}). The cloud
is $\sim$ 60 pc in size with a total gas mass of nearly $4\times 10^5$
M$_{\odot}$, of which $\sim$ 40\% is in a boundary layer referred to as 
the high-density layer (HDL) (Lada et al.~\cite{L78}). The mean number 
density in the cloud is around 40 cm$^{-3}$ implying a very clumpy medium
in the cloud. A number of high-mass star-forming regions are found 
in the HDL which runs parallel to the edge of the 
W4 \ion{H}{ii} region and is likely to be formed by the expansion of the 
\ion{H}{ii} region and the stellar winds of the W4 OB association. Recently,
Moore et al.~(\cite{Metal07}) surveyed W3 and found that
the fraction of the total cloud mass in dense clumps is as high
as 0.26 in the HDL and only 0.05 in the diffuse part of the cloud.
These values are consistent with those obtained in our simulations.

Our simulations support the idea that molecular clouds are 
transient objects.
In this view, star formation happens quickly and the 
parental GMCs are short-lived with observations suggesting lifetimes 
of the order of 10 -- 20~Myr (see e.g. Blitz et al.~\cite{Betal07}).
A lower limit of a few Myr is set by the formation of H$_2$ and CO 
from the atomic gas (Bergin et al.~\cite{Betal04}). Although the  
lifetimes of the cloud in our simulations can only give a rough guide
because  the cloud compression into a disk is fastest when 
the external flow is along the magnetic field lines, they agree well with 
observed values. For strong shocks ($M \geq 5$), however, the
evolution seems to be too fast to produce giant molecular clouds with
morphologies as those observed. Nevertheless, strong shocks can still produce 
molecular clouds as the formation of molecules continues within the 
small coherent structures that fragment off the initial cloud.

Our results would also seem to rule out  very weak shocks 
as agents for molecular-cloud formation since  they do not produce
clumped clouds. A deficiency of our model is that it does not include 
any feedback from star formation such as proto-stellar winds and jets, and 
expanding \ion{H}{ii} regions. However, following the sequential 
star formation picture of Elmegreen \& Lada (\cite{EL77}), we argue 
that these processes occur in the boundary layer of the cloud and 
inject sufficient energy into the cloud to produce    
a clumped medium.  We, therefore, conclude that only weak to 
moderate-strength 
shocks are the likely triggers for the formation of GMCs in the ISM. 

\acknowledgements
We thank the anonymous referee for his/her useful comments that 
improved the original manuscript and J. Urquhart for useful discussions. 
SVL thanks PPARC for the financial support.

\end{document}